\documentclass[10pt,conference,letterpaper]{IEEEtran}

\IEEEoverridecommandlockouts

\usepackage{cite}
\usepackage{amsmath,amssymb,amsfonts}
\usepackage{algorithmic}
\usepackage{graphicx}
\usepackage{textcomp}
\usepackage{xcolor}
\usepackage{makecell}
\def\BibTeX{{\rm B\kern-.05em{\sc i\kern-.025em b}\kern-.08em
    T\kern-.1667em\lower.7ex\hbox{E}\kern-.125emX}}

\usepackage{subfigure}
\usepackage{balance}
\usepackage{comment}

\begin{document}

\title{RiSi: Spectro-temporal RAN-agnostic Modulation Identification for OFDMA Signals
}

\author{Daulet Kurmantayev, Dohyun Kwun, Hyoil Kim, {\it Senior Member, IEEE}, and Sung Whan Yoon, {\it Member, IEEE}
	\thanks{Daulet Kurmantayev is with Tele2 Kazakhstan, Almaty 050000, Kazakhstan (e-mail: daulet.kurmantayev@tele2.kz). He performed this research when he was a graduate student at UNIST.}
	\thanks{Dohyun Kwun and Hyoil Kim are with the Department of Electrical Engineering, Ulsan National Institute of Science and Technology (UNIST), Ulsan 44919, Republic of Korea (e-mail: \{kwun1105,hkim\}@unist.ac.kr).}
	\thanks{Sung Whan Yoon is with the Graduate School of AI, UNIST, Ulsan 44919, Republic of Korea (e-mail: shyoon8@unist.ac.kr).}
	\thanks{Hyoil Kim is the corresponding author.}
	\thanks{This work was supported by 
		the National Research Foundation of Korea (NRF) through the Korean Government (MSIT) under Grant 2020R1F1A1071726, and
		the Institute of Information \& communications Technology Planning \& Evaluation (IITP) through the Korean government (MSIT) under Grant (No. RS-2024-00405128, Development of 6G APIs guaranteeing application performance through interactions with cellular systems), (No. 2020-0-01336, Artificial Intelligence Graduate School Program (UNIST)), and (No. 2021-0-02201, Federated Learning for Privacy Preserving Video Caching Networks).
		}
}

\maketitle

\begin{abstract}
RAN-agnostic communications can identify intrinsic features of the unknown signal without any prior knowledge, with which incompatible RANs in the same unlicensed band could achieve better coexistence performance than today's LBT-based coexistence. 
Blind modulation identification is its key building block, which blindly identifies the modulation type of an incompatible signal without any prior knowledge. 
Recent blind modulation identification schemes are built upon deep neural networks, which are limited to single-carrier signal recognition thus not pragmatic for identifying spectro-temporal OFDMA signals whose modulation varies with time and frequency.
Therefore, this paper proposes RiSi, a semantic segmentation neural network designed to work on OFDMA's spectrograms, that employs {\it flattened} convolutions to better identify the grid-like pattern of OFDMA's resource blocks. 
We trained RiSi with a realistic OFDMA dataset including various channel impairments, and achieved the modulation identification accuracy of 86\% on average over four modulation types of BPSK, QPSK, 16-QAM, 64-QAM.
Then, we enhanced the generalization performance of RiSi by applying domain generalization methods while treating varying FFT size or varying CP length as different domains, showing that thus-generalized RiSi can perform reasonably well with unseen data.
%
%
\end{abstract}

\begin{IEEEkeywords}
6G, RAN-agnostic communications, OFDMA, semantic segmentation, domain generalization
\end{IEEEkeywords}

\section{Introduction}
\label{sec:intro}
To pave the way toward 6G communications, there have been many technical proposals, 
among which is the novel concept of {\it RAN-agnostic communications} \cite{6gOulu19,JCN23Lee}. 
Unlike conventional radio access networks (RANs) that  can only recognize compatible co-RAN signals, a RAN-agnostic RAN is capable of identifying intrinsic features (e.g., modulation type) of the signals transmitted by coexisting `incompatible' RANs, 
without any prior knowledge on their protocols nor frame structures.
Exploiting thus-learned features may enable efficient utilization of 
wireless resources, leading to better coexistence among heterogeneous RANs \cite{6gOulu19,6Gvision}.

RAN-agnostic communications would be particularly beneficial for promoting efficient exploitation of unlicensed bands.
LBT(Listen-before-talk) is the de-facto standard of unlicensed coexistence among WLAN, NR-U, IoT/LPWAN, etc., which is based on energy detection of on-going signals to determine the availability of channel access, thus allowing only one transmitter at a time. 
As a result, LBT leads to low spectrum utilization, which is partly why 4G/5G's unlicensed spectrum access via LAA-LTE and NR-U is not prevailing. 
RAN-agnostic RANs, however, can realize much enhanced inter-RAN coordination than LBT, by enabling underlay coexistence (i.e., simultaneous transmissions from multiple RANs while guaranteeing an acceptable interference level to each other) by exploiting the learned features of the coexisting incompatible RANs. 
Accordingly, future RAN-agnostic 6G should be able to more efficiently utilize the vast amount of unlicensed spectrum (including the 1.2GHz bandwidth in the U-NII bands at 6GHz), to cope with ever-increasing spectrum demands.
RAN-agnostic communications leverage blind modulation identification to recognize the modulation type of an unknown interference signal. 
RAN-agnostic nodes can utilize thus-obtained knowledge to determine more advantageous transmission parameters (e.g., modulation type, transmit power, phase rotation in constellation) for better underlay coexistence, 
possibly via distributed reinforcement learning \cite{JCN23Karunanayake}.
%
Furthermore, blind modulation identification requires much less time in rate adaptation 
than popular heuristic algorithms like Minstrel \cite{Minstrel}, 
since the former is realized at the PHY layer with a time scale of several to dozens of milliseconds while the latter is performed at the MAC layer in hundreds of milliseconds.

There have been various deep learning (DL) techniques proposed for blind modulation identification based on CNN/RNN/LSTM \cite{west17deep,o2017introduction,o2018over,liu2017deep,TVT18Meng,Access20Utrilla}.
They, however, only considered the classification of a signal modulated by a specific class, 
thus not instantly applicable to 
today's most popular modulation schemes like OFDM/OFDMA, 
where a signal includes multiple modulation classes varying with spectro-temporal resource blocks.
Although there exist some DL methods proposed for OFDM modulation recognition \cite{Access19Hong,TVT20Zhang,Access21Park}, they are also limited by treating OFDM itself as a modulation class (thus ignoring its internal structure) or by assuming that the modulation type is not time-varying (thus unsuitable for OFDMA).

To tackle OFDMA's resource-block-wise modulation diversity, this paper proposes applying semantic segmentation to the spectrogram of an OFDMA signal.
Spectrogram-based signal representation is promising due to its natural ability to represent the signal's features varying with time and frequency.
Then, semantic segmentation can detect the regions in the spectrogram corresponding to the resource blocks, so that their modulations can be determined individually.

This paper's contributions are three-fold.
\begin{itemize}
\item We propose 
RiSi(\underline{R}AN-agnostic \underline{i}dentification of \underline{S}ignal \underline{i}nterference), a communication-specific semantic segmentation neural network designed for blind modulation identification. The proposed network is evolved from vanilla DeepLab V3+ \cite{chen2018encoderdecoder} by replacing its 2-dimensional convolutions with flattened convolutions to better identify the grid-like pattern of varying modulations within OFDMA's spectrogram, and by 
adopting three-channel inputs with I sample, Q sample, and signal amplitude.
\item We optimize RiSi's performance by adjusting its architectural parameters, investigating their impact on the recognition accuracy to derive the best combination of them. Our extensive experiments have shown that RiSi can achieve identification accuracy up to 92.9\% for BPSK and up to 75.5\% for QAMs, achieving 86\% on average over BPSK, QPSK, 16-QAM, and 64-QAM. In addition, RiSi is shown to outperform other existing DL models in the literature, showing its superiority. 
\item We apply state-of-the-art domain generalization methods (e.g., SWAD \cite{swad}, MLDG \cite{mldg}) to our network to improve its performance in unseen scenarios to make it truly RAN-agnostic. Its generalization performance has been shown to improve by 4-5\% in the train domain and by 1-2\% in the test domain, compared to a naive approach of applying ERM(Empirical Risk Mimization) \cite{lostDG}. 
\end{itemize}

The paper is organized as follows.
Section~\ref{sec:related} overviews related work.
Then, Section~\ref{sec:model} introduces the proposed architecture of RiSi, and
Sections~\ref{sec:evaluation} and \ref{sec:dg} evaluate RiSi's identification and domain generalization performances, respectively.
Finally, the paper concludes with Section~\ref{sec:conclusion}.

\section{Related Work}
\label{sec:related}
\subsection{Neural Networks for Modulation Identification}

There has been huge enhancement in the accuracy and computational 
complexity of modulation classifiers, thanks to the application of DL techniques to raw I/Q samples.
In this regard, various architectures have been proposed, including CNN, RNN, LSTM (and their combinations) \cite{west17deep,o2017introduction,o2018over,liu2017deep,TVT18Meng,Access20Utrilla}, all of which assumed that the received signal in its entirety belongs to a single modulation class, thus not suitable for OFDMA signals that may include varying modulations with resource blocks. 
When treating OFDM itself as a modulation class, OFDM signals experience the loss of internal features 
thus become not classifiable, as shown in \cite{Gaussian}.
Moreover, numerous CNN models proposed for signal constellation recognition require the modulation to remain constant for a long period of time so as to build a correct constellation, which is against the nature of OFDMA.

Among the DL work 
for OFDM modulation recognition, \cite{Access19Hong} is one of the earliest attempts that proposed CNN with I/Q samples as input.
Later on, \cite{TVT20Zhang} developed a CNN-LSTM model to explore OFDM's spatial-temporal properties, while \cite{Access21Park} proposed a CNN model combined with an OFDM parameter estimator that can classify OFDM signals with varying symbol lengths. 
Nevertheless, the aforementioned work have either considered OFDM itself as a modulation class ignoring its internal structure, or assumed that each subcarrier has the same modulation over time, thus still unsuitable for OFDMA.

As a remedy, this paper proposes using spectrogram-based signal representation as input since spectrograms can well visualize signals in the joint time-frequency domains, which suits the problem of OFDMA modulation classification.
Although there exist some studies on using spectrograms for modulation classification like \cite{spec1,spec3,CompCom21Fonseca,WCL22Lin}, they developed CNN for single-carrier signals only, not fully utilizing the time-frequency resolution capability offered by the spectrogram. 


\subsection{Semantic Segmentation for Modulation Recognition}

\cite{changbo2020radar} proposed using semantic segmentation on Choi-Williams spectrograms for modulation classification, showing its ability to handle multiple signals spreading in the frequency domain. 
However, their work only used simple and distinct modulations, since it was designed for radar systems, not for the modulation families typically used in communication systems. 
Moreover, OFDM communication systems feature much more densely-packed subcarriers with fine-grained internal structures, 
making it fundamentally different from the radar signal identification problem. 

In our work, we propose an evolved architecture of semantic segmentation, by replacing typically-used two-dimensional convolutions in DeepLab V3+ \cite{chen2018encoderdecoder} with flattened convolutions, to better address the spectro-temporal resource-block-wise modulation variations in the OFDMA spectrogram.
Moreover, we further fine-tuned the proposed spectrogram-based semantic segmentation architecture, in terms of realistic and effective dataset construction, hyperparameter tuning, etc.

\section{RiSi: Proposed Neural Network Model}
\label{sec:model}
This section first introduces backgrounds of semantic segmentation, and then elaborates RiSi, our proposed neural network for blind modulation identification of OFDMA signals.

\subsection{Semantic Segmentation: Background}

Semantic segmentation is a process of clustering the pixels of an image that belong to the same class \cite{Long_2015_CVPR, ronneberger2015u, Zhao_2017_CVPR,chen2017deeplab,chen2018encoderdecoder}.
Unlike image classification where the label is assigned to an entire image, the goal of semantic segmentation is to label each pixel with its corresponding class.
Deep semantic segmentation is usually built on top of an image classification network, which serves as an encoder to obtain the local features of an image.
Then, a decoder upscales those features back to the original image's size, and feeds them to the softmax layer where classification predictions are made.

A common semantic segmentation architecture is FCN(fully convolutional network) \cite{Long_2015_CVPR}, that utilizes a deconvolution layer to decode features to an input-sized output with pixel-wise classification.
U-Net \cite{ronneberger2015u} adopts multiple deconvolution layers, a so-called contracting path, that successively upsamples a feature to an input-sized output, and employs a skip architecture that connects the encoding and decoding path to achieve precise pixel-wise inference.
DeepLab \cite{chen2017deeplab,chen2018encoderdecoder}, which is a very popular architecture, utilizes ASPP(Atrous Spatial Pyramid Pooling) in the convolutional layers of the encoding path to exploit multi-scale features with varying reception windows of convolutional filters.
%
In the meantime, a widely used loss function in semantic segmentation, which is also adopted in this paper, is the pixel-wise cross-entropy loss averaged across the input space:
\begin{equation}
\mathcal{L}_{\text{CE}} = - \frac{1}{HW} \displaystyle\sum_{i=1}^{W}{ \displaystyle\sum_{j=1}^{H}{ y_{i,j} \log(\hat{y}_{i,j})}},
\label{eq:loss}
\end{equation}
where $y_{i,j}$ and $\hat{y}_{i,j}$ respectively indicate the label and the inference of the pixel at position $(i,j)$ of $(H\times W)$-sized input.

\subsection{Proposed Architecture of RiSi}

The baseline architecture of our proposed model is DeepLab V3+ with a ResNet-18 backbone, a state-of-the-art deep visual model designed for semantic segmentation \cite{chen2018encoderdecoder}.
Although vanilla DeepLab V3+ shows decent performance in vision processing, it is not instantly suitable for the wireless communications domain.
Hence, we propose its evolved version `RiSi' based on the expert knowledge in communications, by replacing its large two-dimensional convolutional filters with flattened convolutions \cite{flattened} and introducing the third image channel with signal amplitude as an extra information.\footnote{Note that the signal amplitude is instantly available from the first two image channels, i.e., I and Q samples, thus not introducing any extra overhead.}

The concept of flattened convolution comes from filter separation. 
When rephrasing the formula in \cite{flattened}, a convolutional layer with filter weights $W_f\in\mathbb{R}^{C\times X\times Y}$ can be separated into a series of 1-dimensional filters, i.e., $\alpha_f\in\mathbb{R}^{C\times 1 \times 1}$, $\beta_f\in\mathbb{R}^{1\times X\times 1}$, and $\gamma_f\in\mathbb{R}^{1\times 1\times Y}$, such as:
\begin{equation}
F*W_f = \big(\big((F*\alpha_f)*\beta_f\big)*\gamma_f\big),
\end{equation}
where $C$ is the number of channels, $(X\times Y)$ is the size of $W_f$, and $F$ is the input feature map.
This paper only assumes spatial orthogonality of the two-dimensional coordinate of a feature map, so we separate a filter into a series of two filters $\beta_{f}\in\mathbb{R}^{C\times X\times 1}$ and $\gamma_{f}\in\mathbb{R}^{C\times 1\times Y}$ for RiSi such that:
\begin{equation}\small
F*W_f = \sum_{c=1}^{C} \Big( \sum_{x'=1}^{X} 
 \big( \sum_{y'=1}^{Y} F(c, x-x', y-y') \beta_f(c, y') \big)
 \gamma_f(c, x') \Big).
\end{equation}

\if false
\begin{equation}
I*W_f = \sum_{x'=1}^{X} \sum_{y'=1}^{Y} \sum_{c=1}^{C} I(c, x-x', y-y')W_f(c,x',y'), 
\end{equation}
where $C$ is the number of channels, $(N\times M)$ is the dimension of the input map and $(X\times Y)$ is the size of filter. 
\textcolor{blue}{Note that in the above, $x$ and $y$ represent two-dimensional coordidnate of the image, and $c$ implies the channel dimension.}
Then, the flattened convolution tries to separate it into a series of 1-dimensional filters:
\begin{equation}
I*W_f = \sum_{x'=1}^{X} \Big( \sum_{y'=1}^{Y} 
 \big( \sum_{c}^{C} I(c, x-x', y-y') \alpha_f(c) \big) 
 \beta_f(y') \Big) \gamma_f(x') ,
\end{equation}
\textcolor{blue}{where $\alpha_f\in\mathbb{R}^{C\times 1 \times 1}$, $\beta_f\in\mathbb{R}^{1\times X\times 1}$, $\gamma_f\in\mathbb{R}^{1\times 1\times Y}$ are the 1-dimensional filters, respective to the three dimensions.
This paper only assumes spatial orthogonality, so we utilize two filters $\beta_{f}\in\mathbb{R}^{C\times X\times 1}$ and $\gamma_{f}\in\mathbb{R}^{C\times 1\times Y}$ for RiSi:}
\begin{equation}
I*W_f = \sum_{c=1}^{C} \Big( \sum_{x'=1}^{X} 
 \big( \sum_{y'=1}^{Y} I(c, x-x', y-y') \beta_f(c, y') \big)
 \gamma_f(c, x') \Big) .
\end{equation}
\fi

The advantages of adopting flattened convolutions are three-fold. 
First, 1-dimensional filters have fewer parameters to optimize, thus speeding up the training process. 
Second, due to the time-frequency orthogonality, spectrograms would have lower ranks than usual images, making the flattened convolution a natural choice in the encoder layers. 
Third, since resource blocks have rectangular shapes by nature, we expect the segmentation output to be also rectangular.
Since rectangular patterns have low ranks, such a grid-like pattern can be achieved by flattened convolutions in the decoder layers.
Therefore, RiSi utilizes DeepLab V3+ with flattened convolutions such that the convolutional filters in the ResNet-18 backbone, ASPP of the encoder part, and the deconvolution filters of the decoder part are replaced by the flattened ones.

In addition to the adoption of flattened convolutions, 
we also attempt to use the third image channel 
(i.e., RGB system's blue channel)
with signal amplitude as additional information, to increase diversity in the input data.
Note that we ruled out utilizing the phase information for the third channel, due to its bounded and cyclic nature, i.e., when the phase is defined in the interval of $[-\pi,\pi]$, there would be discontinuity when calculating the gradient near the two bounds because the phase becomes identical after rotating a multiple of $2 \pi$.

\begin{figure}[t]
\centering
\includegraphics[width=0.99\columnwidth]{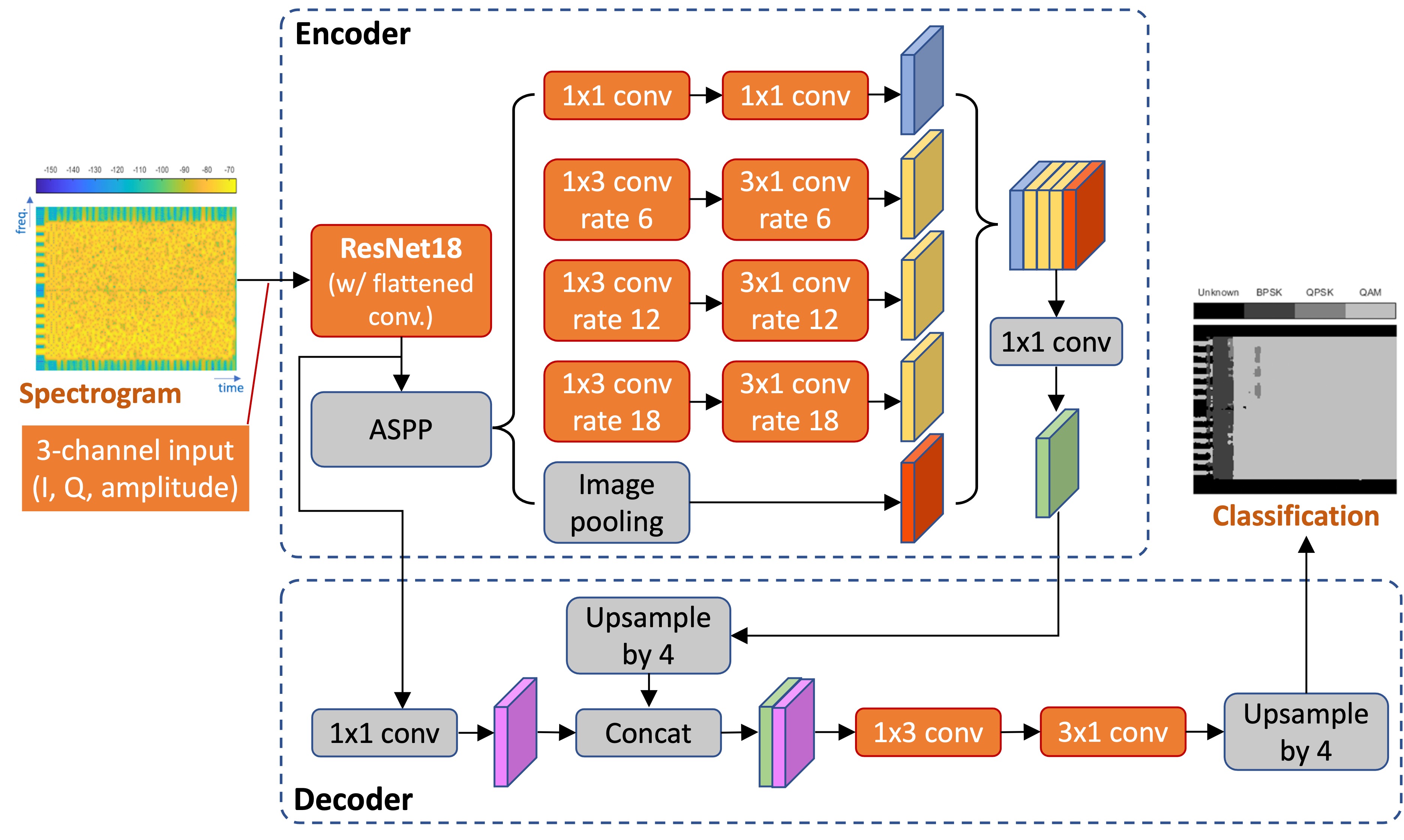}
\vspace{-0.2in}
\caption{Architectural diagram of RiSi (parts in orange are proposed revision)}
\label{RiSi}
\vspace{-0.1in}
\end{figure}

Fig.~\ref{RiSi} illustrates the architecture of RiSi, where the orange-colored parts are our proposal on how to revise the vanilla model. 
In the figure, `rate' means the dilation rate of the ASPP module.
In addition, starting from RiSi with randomly-initialized weights, OFDMA's spectrograms with three channels are fed into RiSi to be trained with the loss in Eq. (\ref{eq:loss}). 

\subsection{Domain-specific Dataset Construction for RiSi}
\label{subsec:dataset_construction}

There exist open source datasets for modulation recognition, such as RadioML 2016.10a \cite{o2016convolutional} and RadioML 2018.01a \cite{o2018over}. 
However, they do not contain OFDMA-modulated signals, so we have developed our own datasets.
Most importantly, to achieve the best-performing RiSi, we need to carefully generate the training dataset according to the intrinsic features of wireless signals using communication-specific knowledge, as follows.
First, we built the dataset consisting not only of synthetic samples but also of over-the-air recordings (of real WLAN signals captured by USRP N210 radios) to ensure our model performs well under real channel conditions. 
To make the synthetic dataset as close to real signals as possible, we considered various channel defects in our model including fading, thermal noise, fractional frequency offset, and clock offset.
Specifically, we have employed the TGn channel delay model-B from the Matlab WLAN Toolbox, with the SNR of 15 dB and a white Gaussian noise, a frequency offset uniformly distributed in $(-312.5, 312.5)$ kHz, and a clock offset uniformly distributed in $(-0.005, 0.005)$. 
Then, a random sequence of bits is OFDMA-modulated with the FFT size of 64, where each OFDMA resource block has a rectangular shape with a randomly-chosen size, and can take a modulation type out of BPSK, QPSK, 16-QAM, 64-QAM. 
We also consider a `no data' class, to address some special cases such as:  when there exists no data at all (e.g., `DC null' in WLAN) or when a frame includes some fields modulated by non-usual schemes (e.g., WLAN header fields).

Practical OFDMA signals may include structures other than resource blocks, such as preambles, header fields, null, guard subcarriers, etc. 
While these structures have different sizes than OFDMA resource blocks, they still occupy rectangular regions in the spectrogram. 
We therefore want to make our system capable of detecting rectangular regions with varying sizes on a given spectrogram.
To do so, in our synthetic dataset, we varied the resource block size between data samples, so that RiSi is trained to detect the structures of different sizes, 
whose efficacy will be shown in Section~\ref{sec:evaluation}.

Thus-synthesized signals are used to obtain STFT(Short-time Fourier transform) based spectrograms with an FFT size of 256, the window length of 256, and the window shift of 8. 
The spectrograms are then transformed into 256x300 RGB PNG images, with the real and complex parts corresponding to the red channel and the green channel, respectively. 
Moreover, we reshaped them to (256x256)-sized images by cropping the exceeding parts of the time axis.
In addition, we also performed an experiment on the effect of using the blue channel to store the amplitude information of the signal, which will be shown in Section~\ref{subsec:channel}.
Note that STFT is obtained by sliding a window function over the signal and applying DFT to the windowed samples. 
More specifically, the STFT matrix of a signal $x(n)$ is defined as:
\begin{equation}
STFT_{x}(f,m) = \sum_{n=mK+1}^{mK+J} x(n)w(n-mK)e^{-j(\frac{2 \pi f}{L})(n-mK)},
\end{equation}
where 
$L$ is the FFT length, 
$w(n)$ is a window function, $J$ is the window function length, and $K$ is the window shift, while a rectangular window is adopted since it will produce the results closest to the OFDM demodulation.

Then, we should also consider: how many samples to generate, and how much portion to allocate for each modulation type.
Most related work have utilized `tens to hundreds' of thousands of signal samples for training, with each sample containing dozens of symbols of a certain modulation type. 
Hence, we initially generated 10k 
samples with four modulation types, BPSK, QPSK, 16-QAM, and 64-QAM, which are evenly distributed in the dataset (i.e., each type is encountered with the same chance).
However, training with the 10k dataset exhibited signs of overfitting in the early stages of training, and the thus-trained network not only showed poor performance but also had trouble with classifying between two QAM classes (see Fig. \ref{cm10k} in Section~\ref{sec:evaluation}). 
Intuitively, 16-QAM and 64-QAM would require more data samples due to their complex signal constellations compared to PSKs. 
Therefore, we appended 20k, 40k, and 90k additional samples consisting only of 16-QAM and 64-QAM data (with an equal chance of appearance) to the original 10k samples, 
producing 30k, 50k, and 100k datasets respectively.
The test results 
have revealed that with an increasing dataset size, overfitting gets much mitigated, and QAMs are more accurately classified, as will be shown in Section~\ref{subsec:dataset_size}.

\section{Identification Performance of RiSi}
\label{sec:evaluation}

This section presents a series of ablation studies based on extensive experiments, to analyze 
1) the impact of replacing 2D convolutions with flattened ones, 
2) the effect of introducing the third channel with amplitude, 
3) the influence of dataset size and organization on the modulation identification performance.
RiSi's real-world performance is also evaluated with the over-the-air WLAN packets.
Finally, RiSi's performance is compared with three existing DL networks in the literature, to show RiSi's superiority in OFDMA identification.

In the sequel, 
we allocated 95\%, 4\%, and 1\% of the dataset for training, validation, and testing, respectively, and 
used the initial learning rate of 0.001 and the learning rate drops by the factor of 0.8 at every epoch.
We also used the ADAM optimizer \cite{adam} in each experiment, since it performed better than SGDM(SGD with Momentum) and RMSProp.
%
The training was conducted by Matlab DL Toolbox running on NVIDIA GeForce RTX 3060 \& 3090 Ti.


\subsection{Effect of Flattened Convolutions}
\label{subsec:flattened}

Using flattened convolution in the decoder serves as a `rectangularity' constraint.
To show its effect, we trained a flattened version of DeepLab V3+ on 10k dataset, while using only two channels with I and Q components for a fair comparison with the two-channel-based vanilla network.
Compared to vanilla DeepLab V3+ achieving 77\% accuracy, the flattened network reaches 78.8\% test accuracy, achieving +1.8\% enhancement.

\begin{figure}[t]
\centering
\includegraphics[width=0.99\columnwidth]{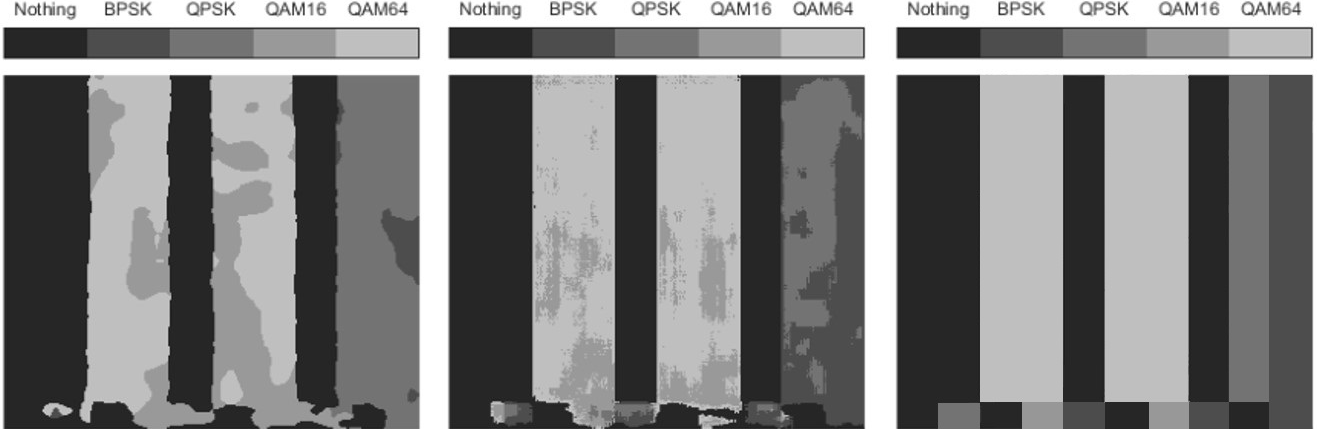}
\vspace{-0.2in}
\caption{Segmentation results: vanilla DeepLab V3+ with 2D convolutions (left), our proposal with 1D convolutions (center), the ground truth (right)}
\label{flatvsbaseline}
\vspace{-0.1in}
\end{figure}

It is also important to visually examine the effect of flattened filters 
in achieving a more rectangular output.
Fig. \ref{flatvsbaseline} presents the outputs of the two networks, showing that the flattened network provides noticeably more rectangular shapes.

\subsection{2-Channel vs. 3-Channel}
\label{subsec:channel}

To reveal the effect of adding the third `amplitude' channel, we used the same signal set to generate the three-channel dataset and the two-channel dataset, and compared their resulting performances.
%
%
In particular, we employed the three-channel version of RiSi (with flattened filters) 
and trained it on the three-channel dataset consisting of 10k samples.
Then, the test accuracy we obtained was 79.8\%, which is another +1\% improvement from 78.8\% reported in Section~\ref{subsec:flattened}. 
Therefore, employing amplitude as an additional input channel indeed has a beneficial effect on the performance.

\subsection{Impact of Dataset Size}
\label{subsec:dataset_size}



\begin{figure}
\centering
  \subfigure[10k dataset]{
    \includegraphics[width = 
	    0.505\columnwidth]{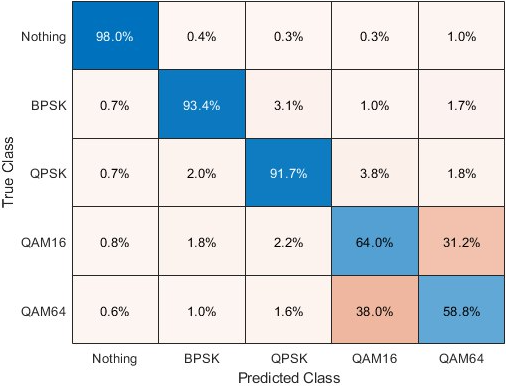}
    \label{cm10k}
    }
  \hspace{-0.1in}
  \subfigure[30k dataset]{
    \includegraphics[width = 
    	0.436\columnwidth]{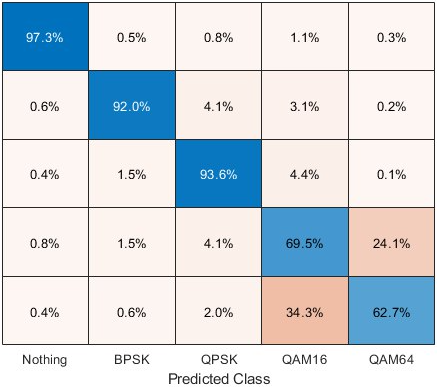}
    \label{cm30k}
    }
  \hspace{-0.07in}
  \subfigure[50k dataset]{
    \includegraphics[width = 
    	0.515\columnwidth]{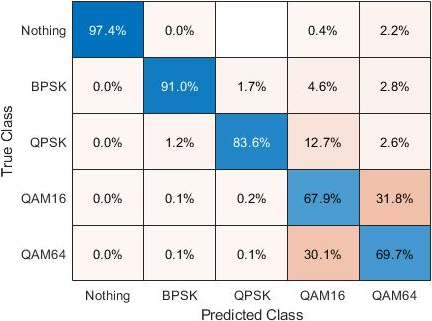}
    \label{cm50k}
    }
  \hspace{-0.11in}
  \subfigure[100k dataset]{
    \includegraphics[width = 
    	0.436\columnwidth]{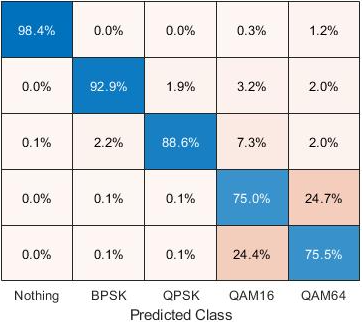}
    \label{cm100k}
    }
\caption{Confusion matrices of RiSi trained on datasets with various sizes (`Nothing' means the `no data' class)}
\label{cm}
\vspace{-0.1in}
\end{figure}

Section~\ref{subsec:dataset_construction} introduced our proposed strategy of constructing the training dataset considering different characteristics among modulation types.
To show the efficacy of the strategy, we varied among 20k, 40k, 90k extra samples (consisting only of 16-QAM and 64-QAM) to be appended to the 10k samples (consisting of all four modulation types) and observed if the performance of the classifier can be improved with the additional data. 
As shown in Fig. \ref{cm}, QAM classification accuracy turns out to be much risen from around 60\% with 10k samples to 75\% with 100k samples. 
The overall system accuracy with the 100k case reaches 86\%, which is +6.2\% enhancement from the 10k case.
Further increase in the dataset size, however, did not show any significant enhancement in the classification accuracy. 


Hence, we conclude that 
the proportion of samples per modulation should be proportional to modulation complexity.

\subsection{Performance with Real-captured Packets}

\begin{figure}[!t]
\centering
  \subfigure[WLAN beacon]{ 
    \includegraphics[width = 0.68\columnwidth]{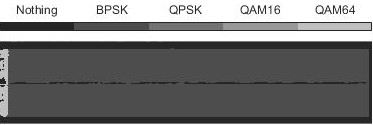}
    \label{real_BPSK}
    }
    \hspace{-0.1in}
  \subfigure[16-QAM data]{ 
    \includegraphics[width = 0.26\columnwidth]{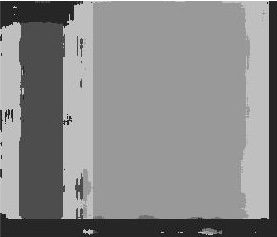}
    \label{real_QAM16}
    }
\vspace{-0.1in}
\caption{A few examples of RiSi's performance with real-captured WLAN packets (vertical axis: frequency, horizontal axis: time)}
\label{real_data}
\vspace{-0.1in}
\end{figure}

To ensure the generalizability of our model to real world signals, we visually tested RiSi with the spectrograms of captured over-the-air WLAN packets.
Fig. \ref{real_data} shows some examples of captured WLAN packets, where Fig. \ref{real_BPSK} presents a BPSK beacon packet's segmentation result from which we can clearly see the preambles and the header at the beginning, as well as a thin horizontal line of the DC null subcarrier.
In addition, Fig. \ref{real_QAM16} shows a high-throughput 16-QAM data packet, from which we observe the BPSK header and the 16-QAM payload correctly identified. 
%
Unfortunately, in the presented cases, STF (the short training field in WLAN's OFDM PLCP preamble) was mis-classified as 64-QAM, which makes some sense since STF has large amplitude variations between subcarriers making it look like 64-QAM.

In conclusion, we observed that our proposed network also works well with real-world signals and can be effectively utilized to visualize the structure of real packets.

\begin{table*}[!t]
\centering
\begin{tabular}{ c|cccccc }
\Xhline{3\arrayrulewidth}
 & \textbf{S-AMR} \cite{spec4} & \textbf{Patched S-AMR} & \textbf{CNN-LSTM} \cite{TVT20Zhang} & \textbf{Patched CNN-LSTM} & \textbf{YOLO} \cite{yolo} & \textbf{RiSi} (Ours) \\
\hline\hline
no data &  100\% & 90.4\% & 100\%   & 93.1\% & 66.0\% & 98.0\% \\
BPSK    &    0\% & 14.1\% & 0\%     & 56.5\% &  4.0\% & 93.4\% \\
QPSK    &    0\% & 0\%    & 0\%     & 18.8\% &  3.0\% & 91.7\% \\
16-QAM  &    0\% & 9.9\%  & 0\%     & 18.4\% &  2.0\% & 64.0\% \\
64-QAM  &    0\% & 80.7\% & 0\%     & 29.3\% &  6.0\% & 58.8\% \\
\hline
Average & 21.2\% & 40.7\% & 20.6\% & 44.4\% & 16.5\% & \textbf{79.8\%} \\
\Xhline{3\arrayrulewidth}
\end{tabular}
\vspace{0.1in}
\caption{Performance Comparison between RiSi and existing Modulation Identification Methods (with 10k dataset)}
\label{tb:perf_comp}
\vspace{-0.24in}
\end{table*}

\subsection{Performance Comparison with Other Existing Methods}

We compared RiSi's performance with three existing alternatives: S-AMR(Spectrogram-based Automatic Modulation Recognition) \cite{spec4}, CNN-LSTM \cite{TVT20Zhang}, and YOLO \cite{yolo}.
S-AMR is based on CNN with 8 convolutional layers to classify a given spectrogram into a single class label. 
For a direct evaluation of S-AMR in OFDMA signals, we assigned the ground truth label as the most frequent modulation type present in the spectrogram.
However, S-AMR should obviously fail due to the lack of ability to find resource-block-wise modulation types, and thus we further extended S-AMR by chunking the spectrogram into multiple patches and assigning the patch-wise label with the most frequent modulation type in the patch (denoted by `Patched S-AMR'). 
Then, when running S-AMR for each patch separately, it is able to learn clear patterns if a single modulation is dominant in each patch.
As another baseline, CNN-LSTM \cite{TVT20Zhang} is originally designed for classifying OFDM signals by stacking three convolutional layers and two LSTM layers, but is still not appropriate to handle OFDMA signals because it does not assume varying modulations along the frequency domain.
Hence, we also extended CNN-LSTM by chunking the spectrogram into multiple patches (denoted by `Patched CNN-LSTM').
In addition, we explored another deep visual model that can localize multiple objects in a given image (i.e., object detection), for which we adopted YOLO, the most well-known object detector.

All the methods (including RiSi) were trained on the 10k dataset, while RiSi is trained with the three-channel data and others are trained with the two-channel data (since using three-channel is our proposal).
For the patched versions, the patch size is set to 32x32.
Note that we have used 10k samples to solely focus on the impact of RiSi's architectural contributions (i.e., semantic segmentation with flattened convolutions and three-channel inputs) against other methods.

For performance comparison, we measured the mean pixel accuracy.
As shown in Table \ref{tb:perf_comp}, S-AMR \cite{spec4} and CNN-LSTM \cite{TVT20Zhang} failed to train OFDMA signals due to the lack of ability to find resource-block-wise identification.
S-AMR can only correctly guess one of many modulations, so it treats the whole spectrogram as a single modulation type thus resulting in the `no data' class. 
CNN-LSTM also fails to correctly classify multiple modulation types along the frequency domain.
Although their patched versions (i.e., Patched S-AMR and Patched CNN-LSTM) show some gains over their original versions, they could not achieve an acceptable performance as well. 
We emphasize that these prior approaches are not tailored to handle OFDMA, thus leading to poor performances in classifying multiple resource blocks simultaneously.
YOLO also presents a poor accuracy in identifying the modulation types, which is as expected considering how they work -- although YOLO naturally identifies rectangular-shaped resource blocks via bounding boxes, its training process is severely hindered by a huge number of closely-packed resource blocks.
On the contrary, RiSi shows an outstanding performance with significant margins against the prior methods.
This clearly signifies the necessity and efficacy of applying semantic segmentation to blind identification of OFDMA signals.\footnote{Note that RiSi's fundamental performance is 86\%, not 79.8\%, which is achieved by our proposed dataset construction scheme in Section~\ref{subsec:dataset_size}.}

\section{Domain Generalization Performance of RiSi}
\label{sec:dg}
Until now, our evaluations have considered a fixed FFT size of 64 and a cyclic prefix (CP) length of 8.
RAN-agnostic communication systems, however, may be involved with numerous parameters 
that may vary according to protocol configurations and operation environments.
Therefore, we need to make RiSi perform satisfactorily even with diverse signal parameter values that may not be seen during training.
Unfortunately, this cannot be fulfilled by simply employing a larger dataset with more parametric variations due to the huge diversity of possible parameter values. 
In this regard, this section investigates how to promote the ability of RiSi in extrapolating the signals unseen at the training stage. 
Specifically, we address the issue as a `domain generalization (DG)' problem by treating varying parameters as different domains.

\subsection{Primer on Domain Generalization Techniques}

In the AI community,
DG tasks aim to train models on `train domains' for generalizing well on `test domains' that have not been encountered.
Among the existing DG approaches, we hereby focus on model- and task-agnostic DG methods such as Empirical Risk Minimization (ERM) \cite{lostDG}, Meta-Learning Domain Generalization (MLDG) \cite{mldg}, and Stochastic Weight Averaging Densely (SWAD) \cite{swad}, which can be easily applied to our task. 
ERM, which simply minimizes the average loss across train domains, has recently been confirmed to achieve competitive performance for DG tasks \cite{lostDG}.
MLDG utilizes the meta-learning strategy for quickly adapting from train domains to test domains \cite{mldg}.
SWAD searches the flatter minima of model weights by taking an ensemble of models, which has been reported to enhance generalization \cite{swad}.
Accordingly, we combine RiSi with the three DG methods as follows.
\textbf{RiSi-ERM} is based on ERM, where RiSi is trained to minimize the averaged loss of various communication parameters in train domains.
\textbf{RiSi-MLDG} meta-trains RiSi by following MLDG. Thus, we randomly split 
the various communication parameters in the train domains into `meta-train domain' and `meta-validation domain' for learning a quick adaptation from the meta-train to the meta-validation split.
\textbf{RiSi-SWAD} takes the ensemble of weights during the stochastic gradient descent, as SWAD works. 
RiSi-SWAD simply trains RiSi through the ensemble process of SWAD.




\subsection{RiSi's Domain Generalization Performance}


Spectrograms of OFDMA signals can largely deviate with varying subcarrier spacing or the CP length, so we hereby consider `varying FFT sizes' and `varying CP lengths' cases.


\begin{figure}[!t]
\centering
  \subfigure[varying FFT size]{
    \includegraphics[width=0.48\columnwidth]{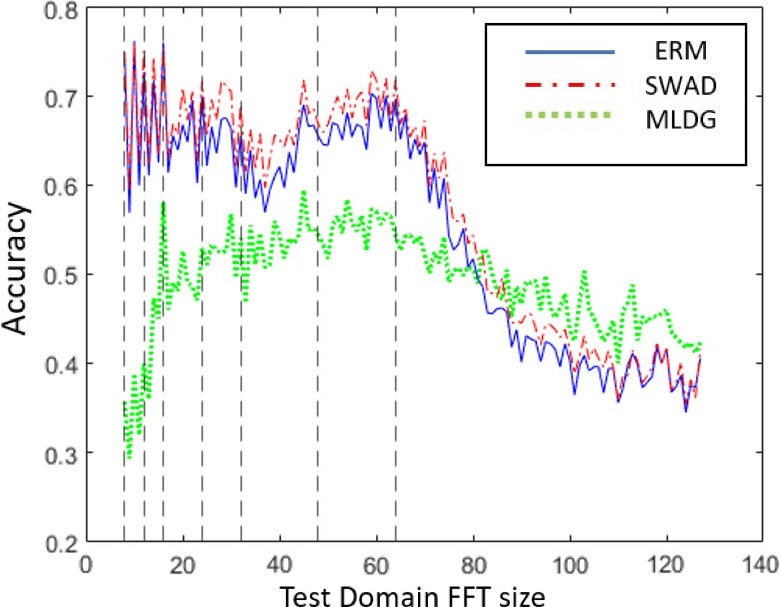}
    \label{fft_dg}
    }
  \hspace{-0.084in}    
  \subfigure[varying CP length]{
    \includegraphics[width=0.464\columnwidth]{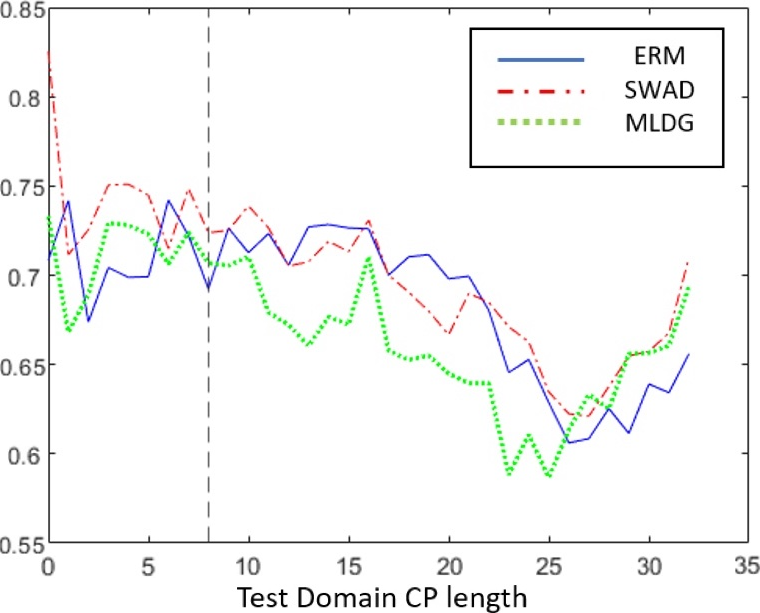}
    \label{cp_dg}
    }
\caption{Evaluation for varying FFT size and CP length. In (a), the vertical dashed lines imply the FFT sizes in the train domain. In (b), the lefthand side of the vertical dashed line construct the train domain.}
\label{fft_cp_dg}
\vspace{-0.1in}
\end{figure}


\textbf{Varying FFT sizes:} 
We first assume a fixed CP length of 8 to focus on varying `FFT size' domains. 
In our evaluation, the train domains are with FFT sizes of $\{8, 12, 16, 24, 32, 48, 64\}$ while the larger sizes of up to 128 are for the test domains.
For all cases, an initial learning rate starts from 0.002.
For RiSi-ERM, the learning rate is exponentially decaying with a factor of 0.92. 
RiSi-SWAD fixes its learning rate to 0.001 after 10 epochs while using a tolerance rate of 1.1.
RiSi-MLDG uses one meta-validation domain at each iteration.
%
As shown in Fig. \ref{fft_dg}, RiSi-SWAD is always performing better than RiSi-ERM, while RiSi-MLDG's performance is not promising except for the FFT sizes larger than 80.

\if false
To justify the sampled source domains (i.e., `sparse') for training instead of using all available domains of $8,9,\ldots,64$ (i.e., `dense'), we trained \textcolor{blue}{RiSi-ERM} and \textcolor{blue}{RiSi-SWAD} for sparse and dense cases, respectively, as shown in Fig. \ref{dg2}.
\textcolor{blue}{Although} dense training does achieve better accuracy than sparse training, the gap is smaller for RiSi-SWAD, and the sparse case performs better out-of-domain with FFT sizes larger than 64 in both SWAD and ERM cases.
Moreover, RiSi-SWAD trained on the sparse dataset is performing almost as good as the densely-trained RiSi-ERM in the train domain and has a superior performance in the test domain FFT sizes larger than 64.
We conjecture that the improved generalization with the sparse dataset is due to the model being overwhelmed with all the domains in the dense training set.
\begin{figure}
\centering
  \subfigure[RiSi-ERM: sparse vs. dense]{
    \includegraphics[width = 0.49\columnwidth]{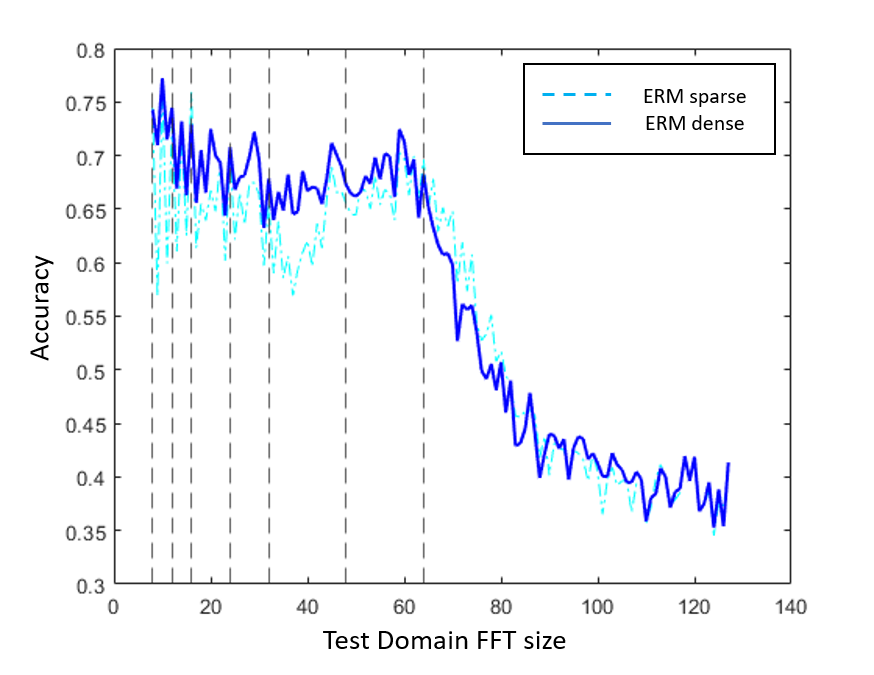}
    \label{10kacc}}
  \hspace{-0.25in}
  \subfigure[RiSi-SWAD: sparse vs. dense]{
    \includegraphics[width = 0.49\columnwidth]{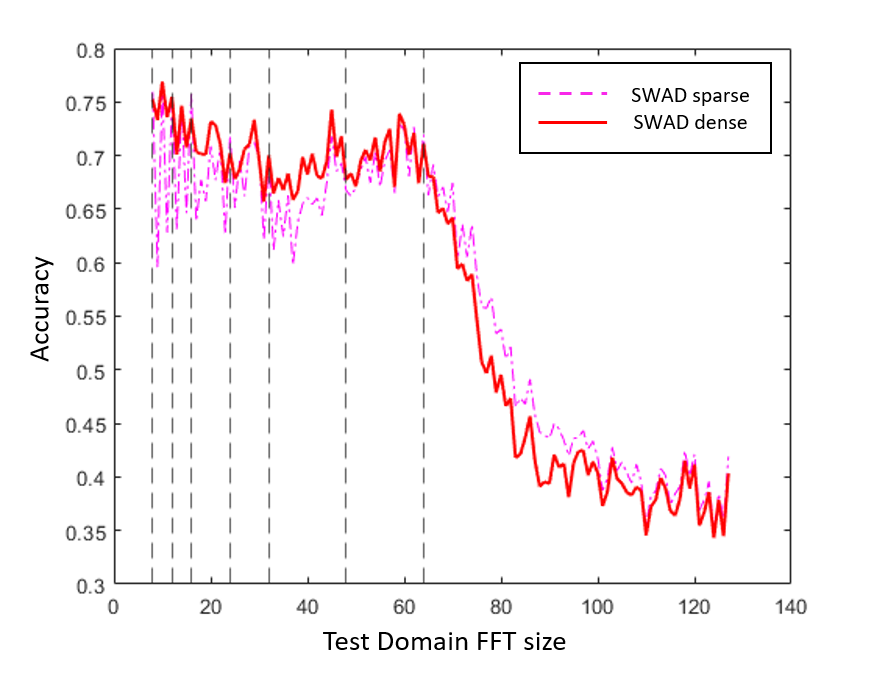}
    \label{10klosssdfbsdf}}
\caption{Training on dense and sparse datasets for RiSi-ERM and RiSi-SWAD}
\label{dg2}
\vspace{-0.1in}
\end{figure}
\fi


\textbf{Varying CP lengths:} Now, we fix the FFT size to 32 while varying the CP length.
The CP length ranges from 0 to 8 for the train domain, while the test domain has a full variety of CP lengths. 
As shown in Fig. \ref{cp_dg}, although the effect of the CP length domain shift is not as substantial as that of FFT size, there is still an improvement after applying the domain generalization methods.

\begin{table}
\centering
\begin{tabular}{ c|ccc }
\Xhline{3\arrayrulewidth}
  & Train domain & Test domain\\
 \hline\hline
 \textbf{RiSi-ERM} (varying FFT)   & 62.2\%    &45.9\%\\
 \textbf{RiSi-MLDG} (varying FFT) & 51.2\% & \textbf{47.7\%}\\
  \textbf{RiSi-SWAD} (varying FFT)&   \textbf{67.8\%}  & \textbf{47.7\%}\\
  \hline
 \textbf{RiSi-ERM} (varying CP)    & 70.9\% & 67.9\%\\
 \textbf{RiSi-MLDG} (varying CP)& 71.2\%  & 65.6\%\\
  \textbf{RiSi-SWAD} (varying CP)&   \textbf{74.4\%}  & \textbf{68.6\%}\\
\Xhline{3\arrayrulewidth}
\end{tabular}
\vspace{0.1in}
\caption{Generalization for unseen FFT sizes and CP lengths}
\label{tb:dg_perf}
\vspace{-0.2in}
\end{table}

Table~\ref{tb:dg_perf} summarizes the obtained mean accuracies for train and test domain performances, in the two training scenarios (i.e., varying FFT sizes and CP sizes). 
We can confirm that DG methods, specifically RiSi-SWAD, allow us to improve test domain accuracy by approximately +1$\sim$2\%, while also improving train domain performance by +4$\sim$5\%, all compared to RiSi-ERM.
RiSi-MLDG outperforms ERM in terms of test domain accuracy in the varying FFT scenario, but its overall performance is poor. 
In conclusion, RiSi-SWAD seems to be the best approach, consistently outperforming other cases. 



\section{Conclusion}
\label{sec:conclusion}
This paper proposed RiSi, a novel neural network architecture for blind modulation identification of OFDMA-modulated signals, built upon semantic segmentation. 
Then, we verified the efficacy of the proposed architecture via extensive evaluations, and 
further applied domain generalization techniques to improve its performance with unseen system parameters. 

Considering the popularity of OFDMA in today's wireless communication standards, we expect our work can be an important stepping stone in realizing our vision of truly RAN-agnostic inter-RAN coexistence. 
As ongoing work, we are currently extending RiSi to make it capable of differentiating mixed interference signals to identify each of them individually.
Also, we are developing a multi-agent reinforcement learning mechanism called RaCE, which can achieve near-optimal performance between coexisting RANs.

\balance		
\bibliographystyle{IEEEtran}
\bibliography{references}

\end{document}